\RequirePackage{xcolor}
\documentclass[journal,twoside,web]{ieeecolor}
\usepackage{tmi}
\usepackage{amsmath,amssymb,amsfonts,centernot}
\usepackage{upgreek}

\usepackage{pifont}
\newcommand{\cmark}{\ding{51}}%
\newcommand{\xmark}{\ding{55}}%

\usepackage[noline,ruled,linesnumbered]{algorithm2e}

\SetCommentSty{mycommfont}

\usepackage{soul}
\usepackage{graphicx}
\usepackage{textcomp}
\def\BibTeX{{\rm B\kern-.05em{\sc i\kern-.025em b}\kern-.08em
    T\kern-.1667em\lower.7ex\hbox{E}\kern-.125emX}}
\usepackage{booktabs, multirow} 
\usepackage{changepage,threeparttable} 
\usepackage{bookmark}
\usepackage{threeparttable}
\usepackage{adjustbox}
\newif\ifgeneratetestwarnings
\generatetestwarningstrue

\newif\ifsuppressclasswarnings
\suppressclasswarningstrue
\ifsuppressclasswarnings
    \usepackage{etoolbox}
    \usepackage{atbegshi}

    \pretocmd{\maketitle}{%
        \newcounter{titlepage}\setcounter{titlepage}{\value{page}}%
        \newcount\oldhbadness\oldhbadness=\hbadness\hbadness=10000
        \newdimen\oldhfuzz\oldhfuzz=\hfuzz\hfuzz=162pt%
        \newdimen\oldvfuzz\oldvfuzz=\vfuzz\vfuzz=22pt%
    }{}{Error.}

    \AtBeginShipout{%
        \ifnumequal{\value{page}}{\value{titlepage}}{%
            \global\hbadness=\oldhbadness%
            \global\hfuzz=\oldhfuzz%
            \global\vfuzz=\oldvfuzz%
        }{}%
    }

    \makeatletter
    \patchcmd{\@evenhead}{%
        \vbox{\color{subsectioncolor}\hrule height1pt width43pc depth0pt}%
    }{%
        \parbox[c][15pt][c]{\textwidth}{\color{subsectioncolor}\hrule height1pt width43pc depth0pt}%
    }{}{Error.}
    \patchcmd{\@oddhead}{%
        \vbox{\color{subsectioncolor}\hrule height1pt width43pc depth0pt}%
    }{%
        \parbox[c][15pt][c]{\textwidth}{\color{subsectioncolor}\hrule height1pt width43pc depth0pt}%
    }{}{Error.}
    \makeatother
\fi
\makeatletter
\let\NAT@parse\undefined
\def\@footnotecolor{red}

\define@key{Hyp}{footnotecolor}{%
 \HyColor@HyperrefColor{#1}\@footnotecolor%
}
\def\@footnotemark{%
    \leavevmode
    \ifhmode\edef\@x@sf{\the\spacefactor}\nobreak\fi
    \stepcounter{Hfootnote}%
    \global\let\Hy@saved@currentHref\@currentHref
    \hyper@makecurrent{Hfootnote}%
    \global\let\Hy@footnote@currentHref\@currentHref
    \global\let\@currentHref\Hy@saved@currentHref
    \hyper@linkstart{footnote}{\Hy@footnote@currentHref}%
    \@makefnmark
    \hyper@linkend
    \ifhmode\spacefactor\@x@sf\fi
    \relax
  }%
\makeatother

\usepackage{hyperref}
\hypersetup{
     colorlinks = true,
     linkcolor = black,
     citecolor= green,
     footnotecolor=black,
}
\usepackage[labeled]{multibib}
\usepackage{cite}

\title{Treatment-aware Diffusion Probabilistic Model for Longitudinal MRI Generation and Diffuse Glioma Growth Prediction}
\author{Qinghui Liu, Elies Fuster-Garcia, Ivar Thokle Hovden, Bradley J MacIntosh, Edvard O.S. Grødem, Petter Brandal, Carles Lopez-Mateu, Donatas Sederevičius, Karoline Skogen, Till Schellhorn, Atle Bjørnerud, and Kyrre Eeg Emblem
\thanks{This work was supported by the European Union's Horizon 2020 Programme: ERC Grant Agreement No. 758657-ImPRESS, Helse Sør-Øst Regional Health Authority grants 2021057, 2017073, the Research Council of Norway grants 325971, 261984, and grant PID2021-127110OA-I00 (PROGRESS) funded by MCIN/AEI/10.13039/501100011033 from Agencia de Investigación de Espana and by ERDF A way of making Europe. (Corresponding authors: Qinghui Liu and Elies Fuster-Garcia.)}
\thanks{Q. Liu, I. T. Hovden, B. J. MacIntosh, E. O.S. Grødem, D. Sederevičius, P. Brandal, K. Skogen, T. Schellhorn, A. Bjørnerud and K. E. Emblem are with the Department of Physics and Computational Radiology, Oslo University Hospital (OUS), Rikshospitalet, 0372 Oslo, Norway (e-mail:qiliu@ous-hf.no; ivarth@student.matnat.uio.no; brad.macintosh@utoronto.ca; edvardgr@uio.no; pebra@ous-hf.no; donsed@ous-hf.no; kaskog@ous-hf.no; UXSCTI@ous-hf.no;  atle.bjornerud@fys.uio.no, kemblem@ous-hf.no). }
\thanks{E. Fuster-Garcia and C. Lopez-Mateu are with the Biomedical Data Science Laboratory, Instituto Universitario de Tecnologías de la Información y Comunicaciones, Universitat Politècnica de València, 46022 Valencia, Spain (e-mail:elfusgar@upv.es, clopmat@upv.es).}
}
\begin{document}
\long\def\/*#1*/{}  
\maketitle

\begin{abstract}
Diffuse gliomas are malignant brain tumors that grow widespread through the brain. The complex interactions between neoplastic cells and normal tissue, as well as the treatment-induced changes often encountered, make glioma tumor growth modeling challenging. In this paper, we present a novel end-to-end network capable of future predictions of tumor masks and multi-parametric magnetic resonance images (MRI) of how the tumor will look at any future time points for different treatment plans. Our approach is based on cutting-edge diffusion probabilistic models and deep-segmentation neural networks. We included sequential multi-parametric MRI and treatment information as conditioning inputs to guide the generative diffusion process as well as a joint segmentation process. This allows for tumor growth estimates and realistic MRI generation at any given treatment and time point. We trained the model using real-world postoperative longitudinal MRI data with glioma tumor growth trajectories represented as tumor segmentation maps over time. The model demonstrates promising performance across various tasks, including generating high-quality multi-parametric MRI with tumor masks, performing time-series tumor segmentations, and providing uncertainty estimates. Combined with the treatment-aware generated MRI, the tumor growth predictions with uncertainty estimates can provide useful information for clinical decision-making.

\end{abstract}


\section{Introduction}
\label{sec:introduction}
\IEEEPARstart{D}{iffuse} gliomas are the most common central nervous system tumors~\cite{louis20212021}, accounting for 80\% of all malignant brain tumors, 30\% of all primary brain tumors, and a major cause of death from primary brain tumors~\cite{schwartzbaum2006}. Despite advances in diagnosis and therapy, the prognosis of diffuse glioma patients, particularly those with glioblastomas, remains poor. Glioblastoma patients have a median survival time of less than 15 months after initial diagnosis~\cite{stupp2009effects, zade2021neuro}. Only very few treatment improvements have been made the last 20 years~\cite{di2023tumor}, and new approaches are highly needed to better understand and estimate glioma tumor growth and treatment response~\cite{henriksen2022high}. Improved growth prediction can provide useful insights for diagnosis, prognosis, personalized medicine, and therapy planning~\cite{soto2024organoid}.

One of the distinguishing features of diffuse glioma growth, particularly glioblastoma, is the highly irregular and unpredictable growth patterns involving multiple tissue types, which make the prediction of diffuse glioma growth notoriously difficult~\cite{petersen2020learning}. Most existing approaches for modeling diffuse glioma growth rely on variants of reaction-diffusion equations that characterize tumor invasion (diffusion term) and proliferation (reaction term)~\cite{clatz2005realistic, harpold2007evolution, hormuth2015predicting}. 
To capture mechanical deformation, the reaction-diffusion model can be coupled to a linear elasticity model for brain and tumor mechanical properties~\cite{clatz2005realistic, hogea2008image}. Extension of these models coupled to nonlinear elasticity has also been presented~\cite{oden2013selection, wong2015tumor, oden2016toward}, but these models are more difficult to calibrate either because they are computationally expensive or they have too many parameters extracted from the imaging data~\cite{tuncc2021modeling}.

Recently, deep-learning-based approaches have been developed to address the glioma growth prediction problem ~\cite{petersen2019deep, gaw2019integration, elazab2020gp, petersen2021continuous}, by learning growth patterns directly from the MRI data. Gaw et al.~\cite{gaw2019integration} (2019) tried to combine a reaction-diffusion model (proliferation-invasion mechanistic tumor growth model) with a data-driven graph-based semi-supervised learning model to provide spatially resolved tumor cell density predictions on primary glioblastoma patients with pre-operative MRI. The authors of~\cite{elazab2020gp} (2020) explored the possibility of using generative adversarial networks (GAN~\cite{gan2014}) for predicting pre-operative glioma growth on both high-grade and low-grade. Other researchers argue that the growth process is not deterministic (e.g., in~\cite{petersen2019deep, petersen2021continuous}) and introduced a deep probabilistic generative model for learning glioma growth dynamics, i.e., estimating a distribution of possible future changes given previous observations of the same tumor. However, all of these approaches have been constrained to using only image data to predict tumor masks while ignoring the effects of any other factors, such as the administered treatment and changes in it. These interventions clearly influence glioma growth behavior and should be accounted for. Additionally, these models also lack the ability to generate future MR images and associated uncertainty maps. Collectively, these limitations are barriers to clinical translation. 

The key challenge of our work stems from the intrinsic complexity of modeling glioma growth under treatment influence~\cite{bottcher2018modeling}. Gliomas exhibit highly heterogeneous, non-linear, and patient-specific growth patterns, shaped by both intrinsic biological factors and external treatments. This variability is exacerbated by the differential impacts of various treatment regimens on tumor progression~\cite{kaznatcheev2017cancer}, creating a significant hurdle in generalizing the model across diverse patients and treatment protocols.

Our research addresses this challenge by developing a model capable of accounting for these multifaceted treatment effects. We introduce the Treatment-aware Diffusion (TaDiff) probabilistic model, which leverages recent advances in diffusion probabilistic models (DPM)~\cite{sohl2015deep, song2019generative, ho2020denoising} and representation learning~\cite{kingma2013auto, rezende2014stochastic, unet2015}. TaDiff integrates treatment data with sequential MRI scans to predict tumor growth and generate future MRIs, effectively overcoming the limitations of earlier approaches.

\begin{figure*}[!t]
\centerline{\includegraphics[width=\textwidth]{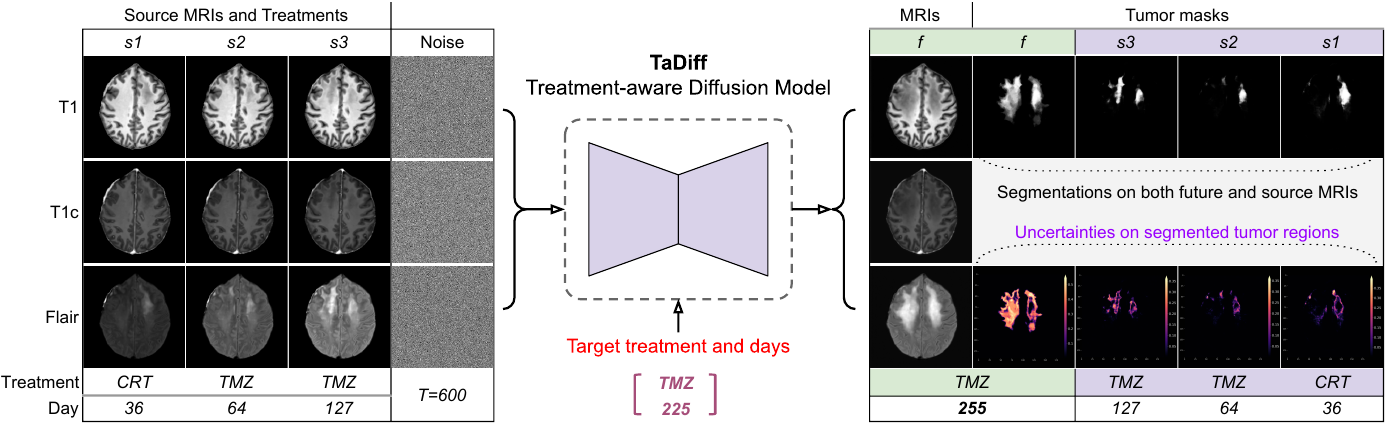}}
\caption{An overview of the TaDiff model (short for Treatment-aware Diffusion Probabilistic model). The goal of our method is to generate a set of synthetic MRIs and tumor progression masks for any given target/future treatment (e.g., TMZ: temozolomide) and time point (e.g., Day: 225) with source sequential MRIs (e.g., \textit{s1}, \textit{s2}, and \textit{s3}) and treatments (e.g., CRT: chemoradiation at Day 36, TMZ at Days 64 and 127). More details are presented in Section~\ref{method}.}
\label{tadiff_1}
\end{figure*}

Specifically, the TaDiff model learns treatment-aware tumor growth dynamics by incorporating a sequence of source MRIs and treatment variables as conditioning inputs. It employs an optimized joint-task learning strategy that combines the strengths of DDPM and tumor segmentation neural networks. This approach enables the generation of future MRIs using a reverse diffusion process for any given treatment and time point, as well as predicting tumor growth through a joint tumor segmentation process. By performing multiple samplings with TaDiff to account for various potential treatment responses (past, present, and future), the model can predict multiple possible tumor growth scenarios and estimate the uncertainty of tumor evolution. An overview of the TaDiff model is illustrated in Figure~\ref{tadiff_1}.

In summary, our contributions are: 
\begin{itemize}
    \item We propose a new conditional diffusion probabilistic model, the Treatment-aware Diffusion (TaDiff) network, capable of performing multiple tasks simultaneously: predicting future multi-parametric MRI scans, estimating tumor segmentation masks for past and future MRIs, and quantifying uncertainty in tumor evolution. Through joint-task learning, TaDiff captures intricate tumor growth patterns by conditioning on both past images and treatment data, enabling the generation of clinically relevant future MRIs and more accurate predictions of tumor progression.
    \item We also introduce a novel dilated longitudinal tumor fusion weighting mechanism to prioritize focusing on crucial peritumoral tissues in glioma progression. This approach strengthens our model's ability to learn spatiotemporal dependencies between tumor evolution and treatment effects, allowing for a more precise representation of glioma dynamics. This leads to more accurate predictions while ensuring stable and efficient training.
    \item Our extensive empirical studies on local and public high-grade glioma datasets confirm the efficacy of our model. It generates longitudinal glioma growth masks and high-quality synthetic multi-parametric MR images using past images and treatment conditions. Moreover, our method offers uncertainty estimates for each prediction, aiding clinical decision-making. 
\end{itemize}

\section{Preliminary}
\label{prel}
The core ideas of diffusion models~\cite{ho2020denoising, ddpm2021}  can be summarized into two key processes: a forward diffusion process, where noise is added to the data, and a reverse diffusion process where a deep neural network tries to separate the noise from the data. 
\subsubsection{Forward Diffusion Process}
The forward process is defined as a Markov Chain in which Gaussian noise is gradually added into a data sample $\mathbf{x}_0$ (e.g. an image) over $T$ successive steps parameterized by a schedule of variances $\{\beta_t \in (0, 1)\}_{t=1}^T$ to produce a sequence of noise samples $\mathbf{x}_1, \dots, \mathbf{x}_T$. It is described by the formulation: 
\begin{equation}  \label{eq:1}
    q(\mathbf{x}_{1:T} \vert \mathbf{x}_0) = \prod^T_{t=1} q(\mathbf{x}_t \vert \mathbf{x}_{t-1}) \approx \mathcal{N}(\mathbf{0} ; \mathbf{I})\;.
\end{equation}
At each step of the forward process, Gaussian noise is added according to
$q(\mathbf{x}_t \vert \mathbf{x}_{t-1}) = \mathcal{N}(\mathbf{x}_t; \sqrt{1 - \beta_t} \mathbf{x}_{t-1}, \beta_t\mathbf{I})$, where the noisy level is controlled by the variance $\beta_t$ (known as the 'diffusion rate'). The data sample gradually loses its distinguishable features as the step $t$ becomes larger. Eventually when $T \to \infty$, $\mathbf{x}_T$ is equivalent to an isotropic Gaussian distribution.
A key property of the forward process is that an arbitrary sample from the Markov chain can be obtained without computing each step in the sequence. Let $\alpha_t = 1 - \beta_t$ and $\bar{\alpha}_t = \prod_{i=1}^t \alpha_i$, we can produce a sample $\mathbf{x}_t \sim q(\mathbf{x}_t \vert \mathbf{x}_0)$ at an timestep $t$ by the formulation. 
\begin{equation} \label{eq:2}
\mathbf{x}_t = \sqrt{\bar{\alpha}_t}\mathbf{x}_0 + \sqrt{1 - \bar{\alpha}_t}\boldsymbol{\epsilon}\;,
\end{equation}
 where $\boldsymbol{\epsilon} \sim \mathcal{N}(\mathbf{0}, \mathbf{I})$. Thus the forward diffusion process can be rewritten/simplified in terms of $\alpha$ as 
 \begin{equation}
     q(\mathbf{x}_t \vert \mathbf{x}_0) = \mathcal{N}(\mathbf{x}_t; \sqrt{\bar{\alpha}_t} \mathbf{x}_0, (1 - \bar{\alpha}_t)\mathbf{I})\;.
 \end{equation}
 
\subsubsection{Reverse Diffusion Process}
The reverse process is also a Markov chain. In this case, a neural network parameterized by $\theta$ predicts the reverse diffusion process at each timestep described as 
\begin{equation}
    p_\theta(\mathbf{x}_{0:T}) = p(\mathbf{x}_T) \prod^T_{t=1} p_\theta(\mathbf{x}_{t-1} \vert \mathbf{x}_t) \;,
\end{equation}
where 
\begin{equation}\label{eq:5}
    p_\theta(\mathbf{x}_{t-1} \vert \mathbf{x}_t) = \mathcal{N}(\mathbf{x}_{t-1}; \boldsymbol{\mu}_\theta(\mathbf{x}_t, t), \boldsymbol{\Sigma}_\theta(\mathbf{x}_t, t))\;.
\end{equation}
It is noteworthy that the reverse conditional probability is tractable when conditioned on $\mathbf{x}_0$:
\begin{equation}
    q(\mathbf{x}_{t-1} \vert \mathbf{x}_t, \mathbf{x}_0) = \mathcal{N}(\mathbf{x}_{t-1}; {\tilde{\boldsymbol{\mu}_t}}(\mathbf{x}_t, \mathbf{x}_0), {\tilde{\beta}_t} \mathbf{I})\;,
\end{equation}
where 
\begin{equation}
   \tilde{\boldsymbol{\mu}}_t\left(\mathbf{x}_t, \mathbf{x}_0\right)=\frac{\sqrt{\bar{\alpha}_{t-1}} \beta_t}{1-\bar{\alpha}_t} \mathbf{x}_0+\frac{\sqrt{\alpha_t}\left(1-\bar{\alpha}_{t-1}\right)}{1-\bar{\alpha}_t} \mathbf{x}_t \;, 
\end{equation}
and 
\begin{equation}
    \quad \tilde{\beta}_t=\frac{1-\bar{\alpha}_{t-1}}{1-\bar{\alpha}_t} \beta_t\;.
\end{equation}
because of $\mathbf{x}_0 = \frac{1}{\sqrt{\bar{\alpha}_t}}(\mathbf{x}_t - \sqrt{1 - \bar{\alpha}_t}\boldsymbol{\epsilon}_t)$ (Eq.~\ref{eq:2}), then 
\begin{equation} \label{eq:9}
    \tilde{\boldsymbol{\mu}}_t = {\frac{1}{\sqrt{\alpha_t}} \Big( \mathbf{x}_t - \frac{1 - \alpha_t}{\sqrt{1 - \bar{\alpha}_t}} \boldsymbol{\epsilon}_t \Big)}\;.
\end{equation}

\subsubsection{Training}
For the reverse diffusion process, a neural network is trained to approximate the conditional probability distributions, i.e., train $\boldsymbol{\mu}_\theta$ to predict $\tilde{\boldsymbol{\mu}}_t$. Because $\mathbf{x}_t$ is available (Eq.~\ref{eq:9}) as input in training time, it is common to predict $\boldsymbol{\epsilon}$ from the input $\mathbf{x}_t$ at time step $t$, thus
\begin{equation}
    \tilde{\boldsymbol{\mu}}_t \approx \boldsymbol{\mu}_\theta(\mathbf{x}_t, t) := {\frac{1}{\sqrt{\alpha_t}} \Big( \mathbf{x}_t - \frac{1 - \alpha_t}{\sqrt{1 - \bar{\alpha}_t}} \tilde{\boldsymbol{\epsilon}}_\theta(\mathbf{x}_t, t) \Big)}\;.
\end{equation}
By letting $\boldsymbol{\Sigma}_\theta(\mathbf{x}_t, t) = {\tilde{\beta}_t} \mathbf{I}$, and letting the forward variances $\beta_t$ to be a sequence of linearly increasing constants from $\beta_1=10^{-4}$ to $\beta_T=0.02$, and some other simplifications in the work~\cite{ho2020denoising}, we can minimize the MSE loss of the noise to train the neural network.
\begin{equation} \label{eq:10}
    \mathbb{E}_{t \sim [1, T], \mathbf{x}_0, \boldsymbol{\epsilon}} \Big[\|\boldsymbol{\epsilon} - \tilde{\boldsymbol{\epsilon}}_\theta(\mathbf{x}_t, t)\|^2 \Big] \;.
\end{equation}
\subsubsection{Inference}
A neural network trained in the reverse diffusion process can be used to generate data. This is achieved by initializing $\mathbf{x}_T \sim \mathcal{N}(\mathbf{0}, \mathbf{1})$ and, in T steps, denoising the image by using
\begin{equation}\label{eq:12}
    \mathbf{x}_{t-1}=\frac{1}{\sqrt{\alpha_t}}\left(\mathbf{x}_t-\frac{1-\alpha_t}{\sqrt{1-\bar{\alpha}_t}}  \tilde{\boldsymbol{\epsilon}}_\theta\left(\mathbf{x}_t, t\right)\right) + \sqrt{\tilde{\beta}_t} \boldsymbol{z}\;.
\end{equation}
where $\mathbf{z} \sim \mathcal{N}(\mathbf{0}, \mathbf{1})$ is new noise added between each denoising step. 

\section{Methods}
\label{method}
The classical DDPM approach requires only $\mathbf{x}_t$ for training, resulting in arbitrary images $\mathbf{x}_0$ when sampling from random noise during inference. However, our goal is not to generate arbitrary images but to generate realistic MRIs and tumor growth maps for any target (future) treatment-day point from a given sequence of source/conditioning images and treatment information. To this end, we propose the treatment-aware diffusion (TaDiff) model for multi-parametric MRI generation and tumor growth prediction on longitudinal data. Our TaDiff model introduces a treatment-aware mechanism for conditioning a diffusion model while also employing a joint learning strategy to segment the tumor and project its future growth during diffusion processes. Figure~\ref{model_arch} illustrates an overview of the TaDiff pipeline.

\subsection{Problem Settings} Let tumor binary masks $\mathbf{M} \in \mathbb{R}^{L \times H\times W\times D} $ be longitudinal 3D tumor volumes with temporal length $L$. The corresponding longitudinal MRI scans $\mathbf{X} \in \mathbb{R}^{L\times C \times H\times W\times D} $ with $C$ channels. In the current study, we consider $C$ = 3 due to the availability of three inputs: T1-weighted (T1), contrast-enhanced T1 (T1C), and fluid-attenuated inversion recovery (FLAIR) images. The corresponding treatment information is represented as $\mathcal{T} = \{{\uptau}_1, \uptau_2,\dots, \uptau_l,\dots,{\uptau}_{L}\}$, indicating the treatment distribution, with the associated treatment days defined as $\mathcal{D} = \{{d}_1, {d}_2, \dots, {d}_l, \dots, {d}_{L}\} \quad \forall \quad d \in \mathbb{N}_0$ and $0 \leq d_{l-1} < d_{l}$. This work considers two treatment types: chemoradiation (CRT) and temozolomide (TMZ), specified as $\uptau \in \{1, 2\} \sim \mathcal{T}$.

We randomly sample a sorted sequence of three scalar indices from available longitudinal exams as conditional sources, i.e. $\mathcal{S} = \{{s}_1, {s}_2, {s}_3\}$, such that $s_i \in [1, \dots, L-1]$ and $s_i \leq s_{i+1}$. Then we sample a scalar index of future (target) sessions from the rest of future exams, that is, $f \in [s_3+1, \dots, L]$. The set of conditional MRIs $\mathbf{X}$ is $\mathbf{X}^{\mathcal{S}} \in \mathbb{R}^{3 \times C \times H\times W\times D}$ and the set of future/target MRIs is $\mathbf{X}^{f} \in \mathbb{R}^{1 \times C \times H\times W\times D}$, correspondingly, we also get the source tumor masks $\mathbf{M}^{\mathcal{S}} \in \mathbb{R}^{3 \times H\times W\times D}$ and the future tumor growth mask $\mathbf{M}^{f} \in \mathbb{R}^{1 \times H\times W\times D}$. In addition, the associated sequence of paired treatment and day variables, that is, $\langle{\mathcal{T}, \mathcal{D}\rangle}^{\mathcal{S} \cup f} = \{\langle{\uptau_{s_1}, d_{s_1}  \rangle}, \langle{\uptau_{s_2}, d_{s_2}  \rangle}, \langle{\uptau_{s_3}, d_{s_3}  \rangle}, \langle{\uptau_{f}, d_{f}  \rangle}\}$.

The objective is to maximize the posteriors, $p\left(\mathbf{M}^f \mid \mathbf{X}^{\mathcal{S}}\right)$, $p\left(\mathbf{X}^f \mid \mathbf{X}^{\mathcal{S}}\right)$, and $p\left(\mathbf{M}^{\mathcal{S}} \mid \mathbf{X}^{\mathcal{S}}\right)$, i.e., to predict future tumor masks, generate the future MRIs, and segment the source tumor masks, given the conditional MRIs $\mathbf{X}^{\mathcal{S}}$ with a sequence of paired treatments and days $\langle{\mathcal{T}, \mathcal{D}\rangle}^{\mathcal{S} \cup f}$. The three posteriors can be combined to form a joint probability
$p\left(\mathbf{M}^{\mathcal{S} \cup f}, \mathbf{X}^{f} \mid \mathbf{X}^{\mathcal{S}}, \langle{\mathcal{T}, \mathcal{D}\rangle}^{\mathcal{S} \cup f}\right)$. 
Due to the computational complexity of the problem, the model training and inference steps are performed using 2D slices sampled along the dimensions $D$ of $\mathbf{X}$ and $\mathbf{M}$, instead of using the 3D imaging volumes, i.e., 
to maximize the posterior 
$p\left(\mathbf{m}^{\mathcal{S} \cup f}, \mathbf{x}^f \mid \mathbf{x}^{\mathcal{S}}, \langle{\mathcal{T}, \mathcal{D}\rangle}^{\mathcal{S} \cup f}\right)$, where $\mathbf{m}^{\mathcal{S} \cup f} \in \mathbb{R}^{4 \times H\times W}  \sim \mathbf{M}^{\mathcal{S} \cup f}$, $\mathbf{x}^{\mathcal{S}} \in \mathbb{R}^{3C \times H\times W} \sim \mathbf{X}^{\mathcal{S}}$, and $\mathbf{x}^f \in \mathbb{R}^{C \times H\times W} \sim \mathbf{X}^f$.

\subsection{Treatment-aware Diffusion Network} \label{tadiff_net}
\begin{figure*}[!t]
\centerline{\includegraphics[width=0.75\textwidth]{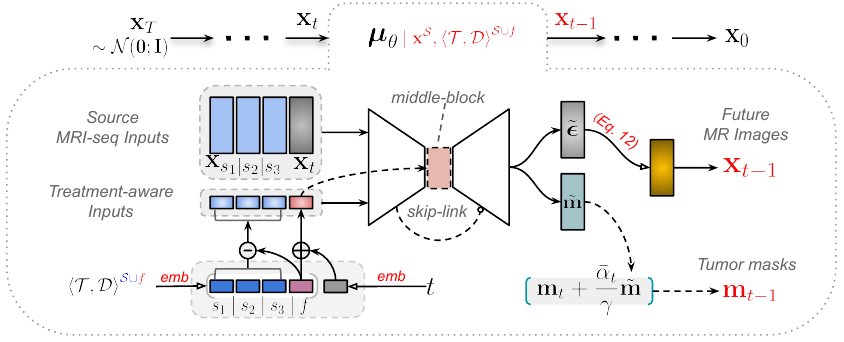}}
\caption{TaDiff model end-to-end pipeline for multi-parametric MRI generation and tumor growth prediction with respect to any given treatment information and target/future time point. In general, the model takes both conditioning source MRI sequences and treatment-aware embeddings as input, outputs target/future MRIs, and predicts tumor masks for both source treatment-day points and the given target treat-day point at the same time. Section~\ref{tadiff_net} presents more details about the network architecture, and algorithms~\ref{alg:train} and~\ref{alg:sample} show details for training and inference with the TaDiff model.}
\label{model_arch}
\end{figure*}

We used a backbone neural network similar to the classical UNet-based diffusion models introduced by Ho et al.~\cite{ddpm2021}. The timesteps of the diffusion process are encoded into the neural network using sinusoidal embedding~\cite{vaswani2017attention} and a single-layer multilayer perceptron (MLP). Following the timesteps embedding idea, we utilize two separate embedding and MLP layers for injecting the paired treatment and day information (i.e., $\langle{\uptau_{s_1},d_{s_1}\rangle}, \langle{\uptau_{s_2},d_{s_2}\rangle}, \langle{\uptau_{s_3},d_{s_3}\rangle},\text{ and},\langle{\uptau_{f},d_{f}\rangle}$) into our network. Specifically, the four pairs of treatment and day are sequentially fed into the two embedding MLP layers and then we sum the learned treatment embedding with its day embedding to get a unique feature vector for each pair of treatment-day variables. This way we get four treatment vectors: three representing source/past treatment information (indicated by $s_1, s_2$ and $s_3$) and one ($f$) representing the target/future treatment information as illustrated in Figure~\ref{model_arch}. 

\begin{algorithm}
    \DontPrintSemicolon
    \SetAlgoLined
    \Repeat{converged}{
    \tcp{Sample conditional and future MRI data. Note that we let $\vert\mathcal{S}\cup f\vert=4$ in this work.}
      $(\mathbf{x}^{\mathcal{S} \cup f},\mathbf{m}^{\mathcal{S} \cup f}, \langle{\mathcal{T}, \mathcal{D}\rangle}^{\mathcal{S} \cup f}) \sim p(\mathbf{X}, \mathbf{M}, \langle{\mathcal{T}, \mathcal{D}\rangle})$\;
      $t \sim \text{Uniform}({1,\dots, T})$\;
      $\boldsymbol{\epsilon} \sim \mathcal{N}(\mathbf{0} ; \mathbf{I})$\;
      \tcp{Eq.~\ref{eq:2}, adding noise into target images}
      $\mathbf{x}_t = \sqrt{\bar{\alpha}_t}\mathbf{x}^f + \sqrt{1-\bar{\alpha}_t}\boldsymbol{\epsilon}$\;
      $(\Tilde{\boldsymbol{\epsilon}}, \Tilde{\mathbf{m}}) = \boldsymbol{\mu}_{\theta}\left(\mathbf{x}^{\mathcal{S}} \cup \mathbf{x}_t, \langle{\mathcal{T}, \mathcal{D}\rangle}^{\mathcal{S} \cup f}, t\right)$\;
      \tcp{Eq.~\ref{seg}, compute the segmentation loss on both source and future tumors}
      $\Tilde{\mathbf{m}}^{\mathcal{S}}, \Tilde{\mathbf{m}}^f =  
 partition(\Tilde{\mathbf{m}})$\;
      $\ell_{seg} = \ell_{dice}(\Tilde{\mathbf{m}}^{\mathcal{S}}, \mathbf{m}^{\mathcal{S}}) + \sqrt{\bar{\alpha}_t}  \ell_{dice}(\Tilde{\mathbf{m}}^f, \mathbf{m}^f)$\;
      \tcp{Computer $\boldsymbol{\omega}$ based on Eq.~\ref{weighting}}
      Take the joint gradient step according to
      $\nabla_\theta\left(\left\|\boldsymbol{\omega}\left(\boldsymbol{\epsilon}-\Tilde{\boldsymbol{\epsilon}}\right)\right\|^2 + \lambda \ell_{seg}\right)$\;
    }
    \caption{TaDiff Training}\label{alg:train}
\end{algorithm}
\begin{algorithm}
    \DontPrintSemicolon
    \SetAlgoLined
    \KwIn{Source images: $\mathbf{x}^{\mathcal{S}}$, source and target (future) treatment-day  $\langle{\mathcal{T}, \mathcal{D}\rangle}^{\mathcal{S} \cup f}$}
    \KwOut{Future MRI: $\mathbf{x}^f$ and tumor masks: $\mathbf{m}^{\mathcal{S} \cup f}$}
    \tcp{$T = 600 $ and $T_m = 10 $ as default}
    Initialize diffusion steps $T$ and mask fusion steps $T_m$\;
    Initialize $\mathbf{x}_{T} \sim \mathcal{N}(\mathbf{0} ; \mathbf{I}) \in \mathbb{R}^{3 \times H\times W}$\;
    Initialize $\mathbf{m}_{T_m} \leftarrow \mathbf{0} \in \mathbb{R}^{4 \times H\times W}$\;
    \tcp{Computer tumor mask fusion factor. }
    $\gamma = \sum^{T_m}_{t=1} \bar{\alpha}_t $\;
    \For{$t\leftarrow T, \dots, 1$}{
        $\boldsymbol{z} \sim \mathcal{N}(\mathbf{0} ; \mathbf{I})$ \textbf{if} $t > 1$ \textbf{else} $\boldsymbol{z} = 0$\;
        $(\Tilde{\boldsymbol{\epsilon}}, \Tilde{\mathbf{m}}) = \boldsymbol{\mu}_{\theta}\left(\mathbf{x}^{\mathcal{S}} \cup \mathbf{x}_t, \langle{\mathcal{T}, \mathcal{D}\rangle}^{\mathcal{S} \cup f}, t\right)$\;
        
        \tcp{Eq.~\ref{eq:12} denoising.}
        $\mathbf{x}_{t-1} = \frac{1}{\sqrt{\alpha_t}}\left(\mathbf{x}_{t}-\frac{1-\alpha_t}{\sqrt{1-\bar{\alpha}_t}} \Tilde{\boldsymbol{\epsilon}}\right)+\sqrt{\tilde{\beta}_t} \boldsymbol{z}$\; 
        \tcp{Fuse the last $T_m$ predicted masks}
        $\mathbf{m}_{t-1} = \mathbf{m}_{t} + \frac{\bar{\alpha}_t}{\gamma}\Tilde{\mathbf{m}}$ \textbf{if} $t \leq T_m$\; 
    }
    \textbf{return} $\mathbf{x}^{f} \leftarrow \mathbf{x}_{0}$, $\mathbf{m}^{\mathcal{S} \cup f} \leftarrow \mathbf{m}_{0} $\;
    \caption{TaDiff Sampling/inference}\label{alg:sample}
\end{algorithm}

Based on experimental observations, we noticed that using the relative (difference) distance between source and target treatment times could help to speed up and stabilize training. As a result, we subtracted the source vectors from the target vectors to get their diff-vectors, which we then concatenate with the summation vector of the target treatment feature and timestep embeddings. We finally used the new concatenated embedding instead of the original timestep embedding to feedforward through the U-Net except for the middle-block part as shown in Figure~\ref{model_arch}. In the middle-block (also known as bottle-neck) of U-Net, we only inject the summation feature between the timestep and target treatment embeddings. Based on our experimental observation, this way provides us a smooth and fast learning process when blending the target-treatment messages into the diffusion timesteps. 

In the input to the model, we concatenated the source multi-parametric MRI (i.e., $\mathbf{x}_{s_1}, \mathbf{x}_{s_2},\text{ and}, \mathbf{x}_{s_3}$) with $\mathbf{x}_t$. Note that $\mathbf{x}_t$ is the target MRI with diffusion noise according to Eq.~\ref{eq:2}. There are two outputs of the model. The first is the diffusion estimated diffusion noise at the time step $t$. This can be used to generate images using Eq.~\ref{eq:12}. The second output is the prediction of the tumor masks from both the source MRIs ($\mathbf{x}^{\mathcal{S}}$) and the noised target MRI ($\mathbf{x}_t$). As a result, rather than being a pure diffusion model, our TaDiff is a hybrid network that combines both denoising diffusion and segmentation tasks, allowing us to generate future MRI and segment longitudinal tumor growth masks simultaneously in an end-to-end manner. We present the extension of the DDPM training and sampling processes in Algorithms~\ref{alg:train} and~\ref{alg:sample} respectively. Note that, the diffusion and segmentation branches share the encoder component. The diffusion process requires 600 steps to generate the target images, whereas the segmentation process utilizes only the final 10 steps. These final steps provide the highest quality input, from which we generate 10 masks. These masks are then weighted and averaged to produce the final segmentation mask. This method ensures that the segmentation masks benefit from the high-quality inputs derived from the latter stages of the diffusion process.

Like~\cite{wolleb2022diffusion}, applying our model with multiple stochastic sampling processes (in this work, we considered 5 samplings), one can obtain pixel-wise uncertainty maps for both the predicted tumor masks and the synthesized MRIs by computing the standard deviation of our model's predictions on the same conditioning inputs.

\subsection{Joint Loss Function}
Because the model predicts a series of tumor masks on both source and future MRIs in addition to the conventional diffusion noise. We hence employ the dice loss~\cite{milletari2016v}, for the segmentation subtask, which is defined as
\begin{equation}\label{eq:dice_loss}
\ell_{dice}(\Tilde{\mathbf{m}}, \mathbf{m})=1 - \frac{2 \left|\Tilde{\mathbf{m}} \mathbf{m}\right|}{\left|\Tilde{\mathbf{m}}\right|+\left|\mathbf{m}\right|}.
\end{equation}
where $\left| \cdot \right|$ denotes the $L_1$ norm. 
Due to the lack of information in $\mathbf{x}_t$ for high $t$, predicting a good mask from the image is unfeasible for the model. We therefore weigh the Dice loss of the future segmentation according to the weight of $\mathbf{x}_0$ in  $\mathbf{x}_t$. The segmentation loss is then defined as
\begin{equation}  \label{seg}
    \ell_{seg} = \ell_{dice}(\Tilde{\mathbf{m}}^{\mathcal{S}}, \mathbf{m}^{\mathcal{S}}) + \sqrt{\bar{\alpha}_t}  \ell_{dice}(\Tilde{\mathbf{m}}^f, \mathbf{m}^f)\;.
\end{equation}
 The first dice loss is used for segmentations ($\Tilde{\mathbf{m}}^{\mathcal{S}}$) of the source images, while the second dice loss is scaled/weighted by the noise level factor $\bar{\alpha}$ used for predicting the tumor masks ($\Tilde{\mathbf{m}}^f$) of noised target images ($\mathbf{x}_t$). 

In addition, we propose a weighting mechanism that enables the model to prioritize and extract representations from the glioma region and its most dynamic and heterogeneous surrounding regions, preventing the model from directly copying the representations of reference source input images and facilitating future MRI construction task learning. To this end, the weighting function is designed as
\begin{equation} \label{weighting}
\begin{split}
\mathbf{\omega} & = \hat{\mathbf{m}}{e}^{-\hat{\mathbf{m}}} \ast \mathbf{f}_{k_l \times k_l} + 1\;.
\end{split}
\end{equation}
where $\hat{\mathbf{m}} = \sum_{i\in \mathcal{S} \cup f} \mathbf{m}_{(i, \cdots)}^{\mathcal{S} \cup f}$, $\ast$ denotes a convolution operation and $\mathbf{f}_{k_l \times k_l}$ is a convolution filter initialized as $0.1$ with a predefined kernel size (default as $k_l = 11$). The weights for the dilated tumor growth region will be in the range $[5.451, 1.886]$, while for the other region, the weights will be $1$. The weight matrix element value reflects the level to which the corresponding spatial feature has changed over longitudinal time points.

Finally, the joint loss function for training the TaDiff model is defined as 
\begin{equation}
\ell_{TaDiff}=\left\|\boldsymbol{\omega}\left(\boldsymbol{\epsilon}-\Tilde{\boldsymbol{\epsilon}}\right)\right\|^2 + \lambda \ell_{seg}\;.
\end{equation}
where $\lambda$ controls the amount of segmentation loss during the joint optimization training process. Guided by our experiments and observation, we set $\lambda = 0.01$ in this work.

\section{Experiments} \label{exp}
\subsection{Dataset and Metric}
\textbf{Local data:} A total of 225 MRI exams~\cite{sailor} were used for this study. There were 23 patients with diffuse high-grade glioma, confirmed by histology, and these patients received treatment at our institution~\cite{sailor2020}. Patients received standard treatment, including surgery, fractionated radiotherapy (approximately four weeks after surgery) with concomitant and/or adjuvant chemotherapy (CRT) with temozolomide (TMZ)~\cite{fuster2022}. There were between 3 and 19 (mean 10) longitudinal MRI scans acquired, consisting of multi-parametric types (used in this study): pre-contrast T1-weighted, post-contrast T1-weighted (T1c), and T2 fluid-attenuated inversion recovery (Flair). Scans were skull-stripped, registered to a common space defined by the T1, and resampled to isotropic $1$ mm resolution. For intensity normalization, we utilized z-score normalization on a per-channel basis (channel means image type, i.e., T1, T1c, Flair, etc). Ground truth segmentations of edema and enhancing tumors were created by experienced neuroradiologists. 
The dataset was manually divided into two parts, a training/validation set, and a test set. The training set included 18 patients and 177 longitudinal scans, while the test set consisted of 5 representative patients from a wide range of age groups (32-65) with a total of 48 longitudinal scans. 

\textbf{External data:} We collected data from 37 additional high-grade glioma patients (18 men, 19 women), totaling 132 MRI exams from the publicly available LUMIERE dataset~\cite{suter2022lumiere}, for further evaluation. All patients underwent surgical resection and temozolomide-based chemo-radiation at Inselspital, the University Hospital of Bern, Switzerland. The mean age at the time of the first resection was 58 years, ranging from 39 to 72 years, with survival times ranging from 224 to 1309 days (mean: 686 days). However, this dataset didn't provide expertly or manually annotated tumor masks. 

\textbf{Metric:} To evaluate the accuracy of the tumor segmentation and growth predictions, we calculated the Dice Similarity Coefficient (DSC) and Relative Volume Difference (RVD) between the ground truth annotations and the estimated tumor maps with an optimized threshold. For evaluating the quality of the generated MRIs, three widely used evaluation metrics were utilized, i.e., the structural similarity index (SSIM)~\cite{wang2004image}, peak signal-to-noise ratio (PSNR), and mean squared error (MSE) between the target and the synthesized multi-parametric MRI. While, on the external test dataset, we only evaluate our model's performance on future MRI generation, assessed using the SSIM, PSNR, and MSE. 

In a clinical setting, the generation of high-quality MRI images significantly improves the accuracy of diagnosing and monitoring conditions such as tumor growth. Higher SSIM or PSNR values indicate greater structural similarity and improved image clarity. Similarly, higher DSC scores and lower RVD values reflect more accurate tumor mask predictions, leading to enhanced tumor growth quantification. These improvements enable precise assessments of tumor boundaries, size, and progression, thereby supporting more informed clinical decisions.

\subsection{Implementation and Training Details}
The proposed method was implemented with PyTorch~\cite{paszke2017automatic}. The model used a 2D U-Net-like structure that consists of an encoder and decoder with skip connections and 64, 128, 256, and 512 channels for each stage. For the diffusion module, we modified the network designed in DDPM~\cite{ddpm2021} into our proposed treatment-aware conditioning model. We employed the Adam optimizer algorithm with an initial learning rate $2.5 \times 10^{-4}$ with a warm-up ($1000$ steps) and a cosine decay scheduler to train the model for 5 million iterations. We used a batch size of $32$ and a gradient accumulation step of $2$. The images were cropped to a patch size of $192 \times 192$. The training took 350 GPU hours using a single Nvidia V100 32GB GPU. 

Note that during the training phase, our model is trained to predict three scenarios: future (probability: 0.5), middle (probability: 0.3), and past (probability: 0.2) MRI exams. This means each MRI slice can serve as either an input or an output, not just future slices as outputs. This approach significantly increases the number of training combinations. For example, with only 10 MRI slices, using only the past 3 to predict the fourth would yield just 7 training examples. However, by allowing each MRI to be an output and selecting 1, 2, or 3 MRI images from the remaining 9 as inputs (allowing repetition), we achieve a total of 8,190 possible training combinations. While time points and slices are likely dependent on one another, this method functions similarly to data augmentation, effectively maximizing the use of our limited data. Moreover, the diffusion training process, which adds random noise, enhances the robustness and effectiveness of the model by exposing it to varied scenarios during training. These strategies collectively improve the model's reliability in predicting future outcomes despite the limited dataset size. During the subsequent testing and inference, we utilize the most recent 3 historical MRI data to predict the nearest future frame. If fewer historical sequences are available, the most recent MRI is duplicated to meet the requirement of three input sources.

\subsection{Test Results}
\begin{figure*}[!htp]
\centerline{\includegraphics[width=0.825\textwidth]{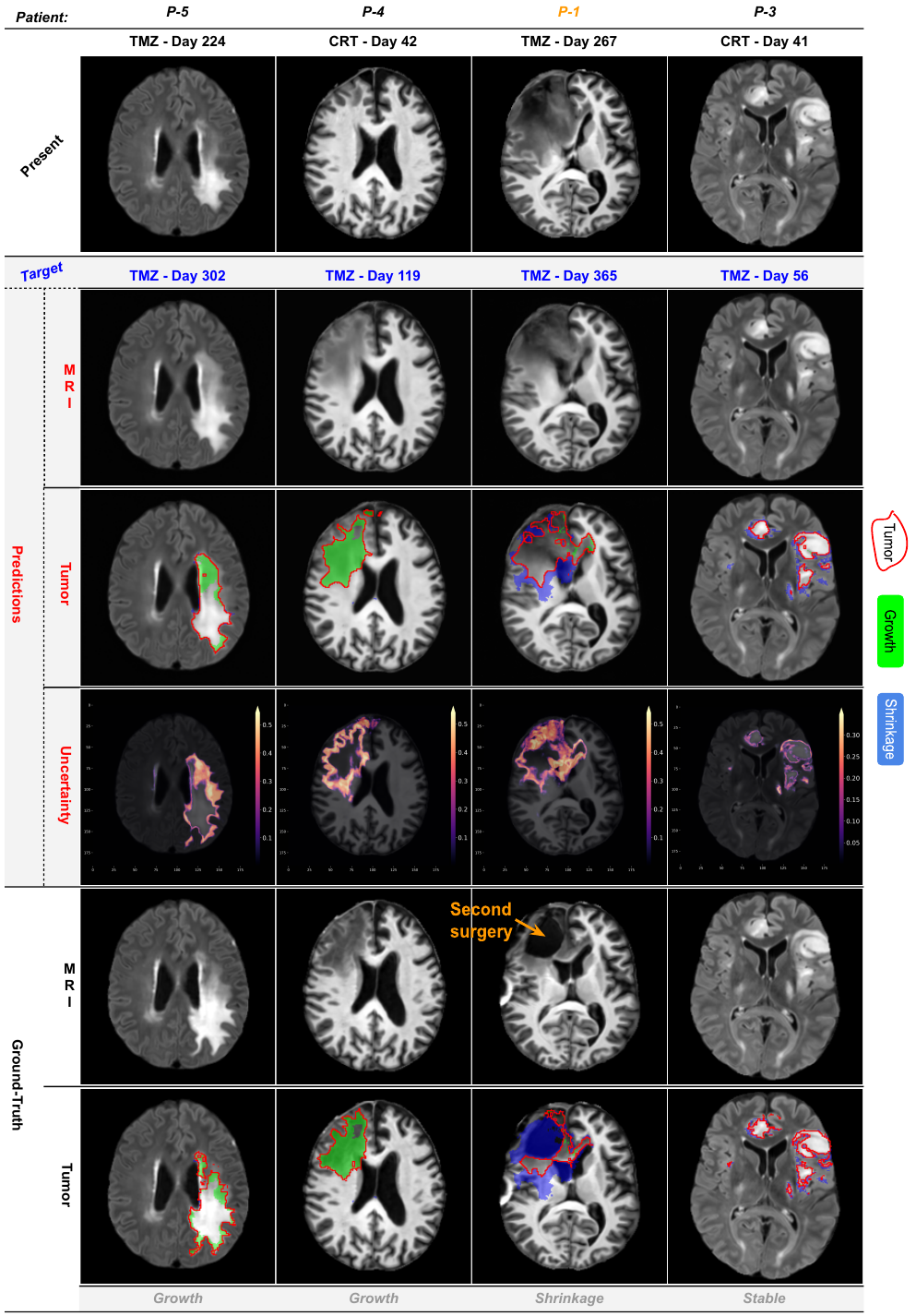}}
\caption{Examples of qualitative predictions on the local test cases. From top to bottom, the Present (source) MRIs with specific treatment-day traces, the Target (future) treatment-day, the Predictions (including generated MRIs, tumor masks, and uncertainty maps), and the Ground truth. The method can model both stable and progressive tumors with growth and shrinkage in different spots. Note that P-1 had a second surgery treatment which was beyond the range of treatment types our model was designed to handle.}
\label{results_growth}
\end{figure*}
\begin{table*}[!htp]\centering
\begin{adjustbox}{max width=\textwidth}
\begin{threeparttable}
\caption{Results on Each Test Patient.}\label{tab: patient}
\begin{tabular}{ccccccccccc}\toprule
\multirow{2}{*}{Patient ID} &\multirow{2}{*}{Age} &\multirow{2}{*}{Survival-month} &\multicolumn{3}{c}{Multi-parametric MRI Generation} &\multicolumn{2}{c}{Future Tumor Prediction} &\multicolumn{2}{c}{Source Tumor Segmentation} \\\cmidrule{4-10}
& & &SSIM $\uparrow$ &PSNR $\uparrow$  &MSE $\downarrow$ &DSC $\uparrow$  &RVD $\downarrow$ &DSC $\uparrow$  &RVD $\downarrow$  \\\cmidrule{1-10}
$^a$ P-1 &32 &84 & $0.886 \pm 0.06$ & $28.141 \pm 1.30$ & $0.155 \pm 0.05$ &$0.691 \pm0.16$ &$0.139 \pm0.69$ &$0.828 \pm0.09 $&$0.087 \pm0.18$  \\
$^b$ P-2 &44 &19 & $0.926 \pm0.04$ &$28.479 \pm1.76$ &$0.138 \pm0.05$ &$0.578 \pm0.17$ &$0.415 \pm1.61$ &$0.818 \pm0.08$ &$0.053 \pm0.19$  \\
P-3 &53 &11 & $0.943 \pm0.01$ &$27.568 \pm0.84$ &$0.212 \pm0.05$ &$0.812 \pm0.06$ &$-0.191 \pm0.10$ &$0.893 \pm0.04$ &$-0.020 \pm0.08$ \\
P-4 &64&13 & $0.903 \pm0.03$ &$27.668 \pm0.84$ &$0.197 \pm0.04$ &$0.757 \pm0.16$ &$-0.187 \pm0.24$ &$0.849 \pm0.16$ &$-0.036 \pm0.24$ \\
P-5 &65 &40 & $0.936 \pm0.03$ &$27.975 \pm1.24$ &$0.166 \pm0.05$ &$0.755 \pm0.10$ &$-0.058 \pm0.31$ &$0.857 \pm0.06$ &$0.033 \pm0.25$ \\ \midrule
Average &52 & 33 & $0.919 \pm0.03$ &$27.966 \pm1.20$ &$0.174 \pm0.05$ &$0.719 \pm0.13$ &$0.024 \pm0.59$ &$0.849 \pm0.09$ &$0.023 \pm0.19$ \\
\bottomrule
\end{tabular}
\begin{tablenotes}
          \footnotesize   
          \item[a] P-1 underwent a second surgery treatment to remove glioblastomas between days 267 and 365. 
          \item[b] P-2 was diagnosed with secondary glioblastomas; such patients were fairly rare (2 cases) in our training dataset.
\end{tablenotes}
\end{threeparttable}
\end{adjustbox}
\end{table*}
\begin{table*}[!htp]\centering
\caption{Quantitative Results w.r.t Treatment Days.}\label{tab: treat}
\scriptsize
\begin{tabular}{lcccccccc}\toprule
\multirow{2}{*}{Treatment Days} &\multicolumn{3}{c}{Multi-parametric MRI Generation} &\multicolumn{2}{c}{Future Tumor Prediction} &\multicolumn{2}{c}{Source Tumor Segmentation} \\\cmidrule{2-8}
&SSIM $\uparrow$ &PSNR $\uparrow$  &MSE $\downarrow$ &DSC $\uparrow$  &RVD $\downarrow$ &DSC $\uparrow$  &RVD $\downarrow$ \\\cmidrule{1-8}
$0 \text{ -- } 50$ & $0.928 \pm0.02$ &$27.924 \pm1.19$ &$0.182 \pm0.06$ &$0.759 \pm0.12$ &$-0.133 \pm0.27 $&$0.858 \pm0.09$ &$-0.012 \pm0.17$ \\
$51 \text{ -- } 220$ & $0.896 \pm0.07$ &$27.984 \pm1.36$ &$0.175 \pm0.06$ &$0.718 \pm0.16$ &$0.113 \pm0.71$ &$0.863 \pm0.08$ &$0.067 \pm0.19$ \\
$221 \text{ -- } 365$ &$0.877 \pm0.06$ &$28.096 \pm1.38$ &$0.163 \pm0.05$ &$0.604 \pm0.20$ &$0.539 \pm1.68$ &$0.843 \pm0.10$ &$0.104 \pm0.39$ \\
$366 \text{ -- } 720$ & $0.926 \pm0.03$ &$28.106 \pm1.33$ &$0.154 \pm0.05$ &$0.697 \pm0.14$ &$-0.024 \pm0.38$ &$0.831 \pm0.08$ &$0.059 \pm0.14$ \\
$721+$ & $0.941 \pm0.02$ &$28.106 \pm1.36 $&$0.149 \pm0.05$ &$0.760 \pm0.09$ &$-0.012 \pm0.27$ &$0.830 \pm0.06$ &$0.039 \pm0.12$  \\
\bottomrule
\end{tabular}
\end{table*}
\begin{figure}[!htp]
\centerline{\includegraphics[width=\columnwidth ]{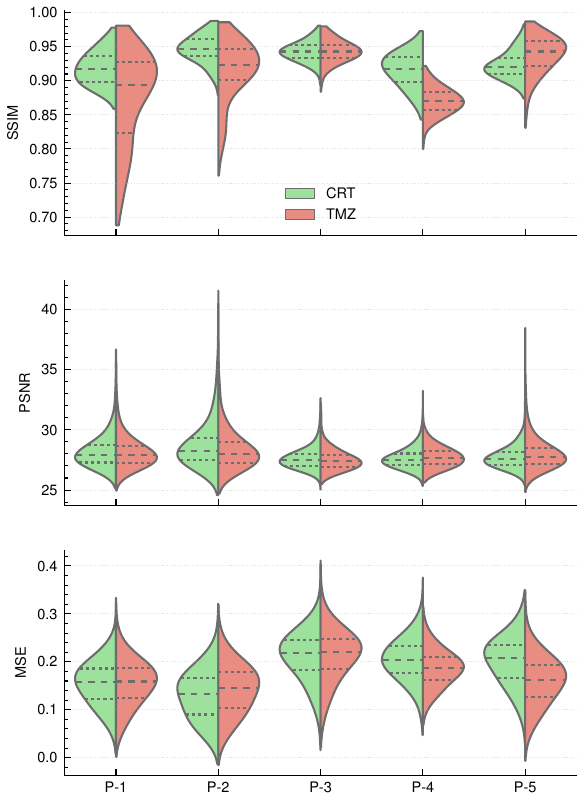}}
\caption{The three split violin plots compare MRI generation metric's distributions of each treatment (CRT and TMZ) overall patients. Dashed lines represent the quartiles for each group. Notice that TMZ has a long-tail distribution below the first quartile for patients P-1 and P-2 with respect to SSIM. The main reason is that P-1 was given a second surgery treatment during the TMZ period, and P-2 was diagnosed with a glioma classified as secondary glioblastoma.}
\label{violin_plot}
\end{figure}
\begin{figure}
\centerline{\includegraphics[width=\columnwidth]{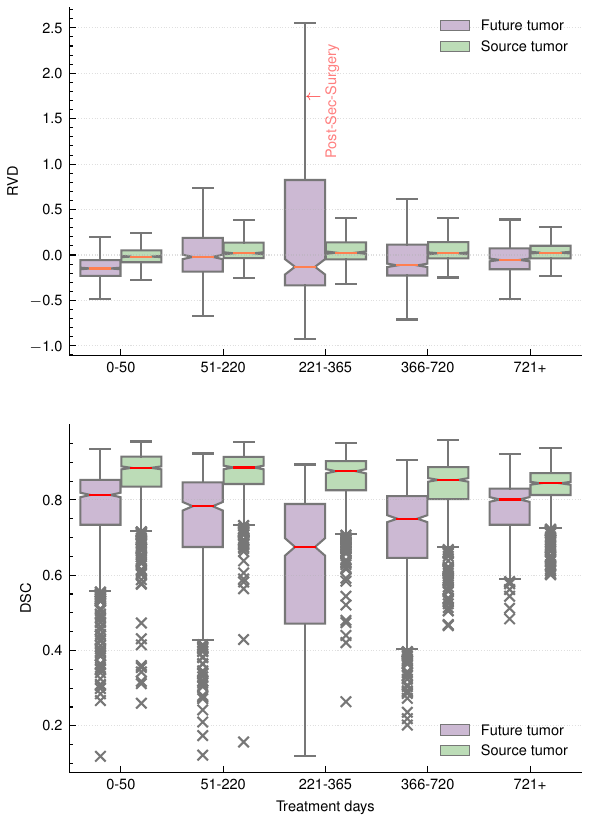}}
\caption{The two box plots show the RVD and DSC distributions for predicted future tumors and source tumors across different treatment day ranges. Note that the model's performance declined significantly with increased variability in the 221-365 day range, this is because one case (P-1) experienced unusual and rapid glioma growth and underwent a second surgery treatment during this period.}
\label{box_plot}
\end{figure}
\begin{figure*}[!htp]
    \centerline{\includegraphics[width=\textwidth ]{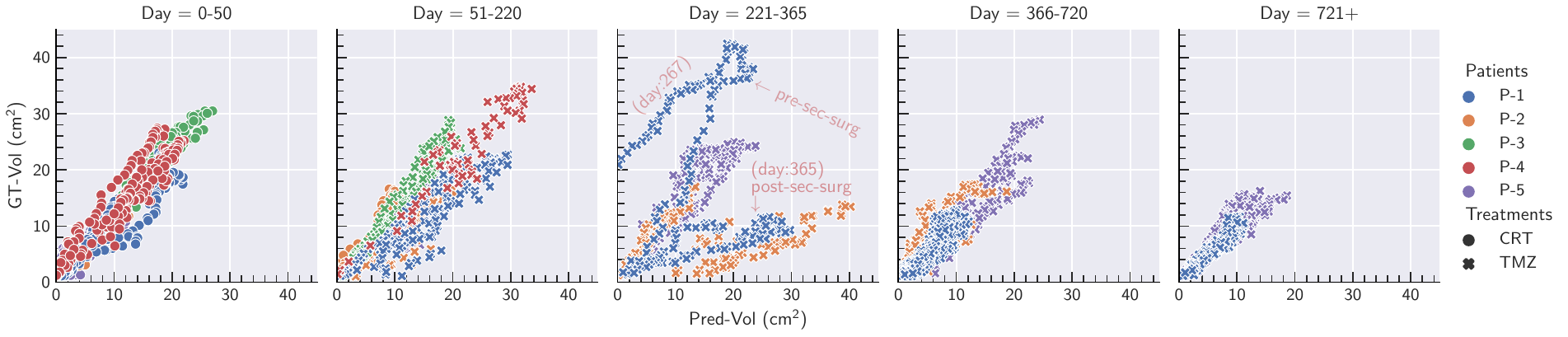}}
    \caption{Tumor growth prediction evaluation in terms of slice-based volume ($\text{cm}^2$) estimation for future treatment and days. This figure shows the detailed relations between GT-Vol and Pred-Vol on each of the day ranges, test patients, and treatment types. Each relational plot corresponds to a target day range, and the patients are differentiated by colors, with different markers indicating the treatment type. Notably, we did not filter out cases involving second surgeries (pre and post) in these plots; instead, we highlighted them with annotations in the day=221–365 plot.}%
    \label{fig:rel_plot}
\end{figure*}
\begin{figure}
    \centerline{\includegraphics[width=0.95\columnwidth ]{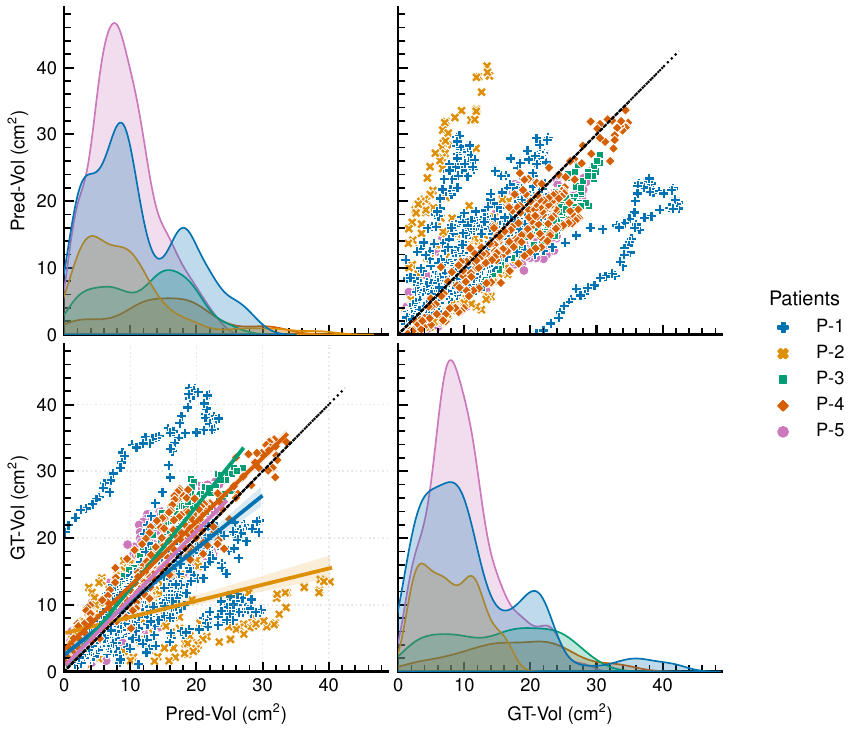}}
    \caption{Tumor growth prediction evaluation on each patient with a paired plot between GT-Vol and Pred-Vol. The plot's identity line indicates the Pred-Vol should ideally equal GT-Vol, and the two distributions of GT-Vol and Pred-Vol reveal their similarities.}%
    \label{fig:pair_plot1}
\end{figure}
\begin{figure}
    \centerline{\includegraphics[width=\columnwidth ]{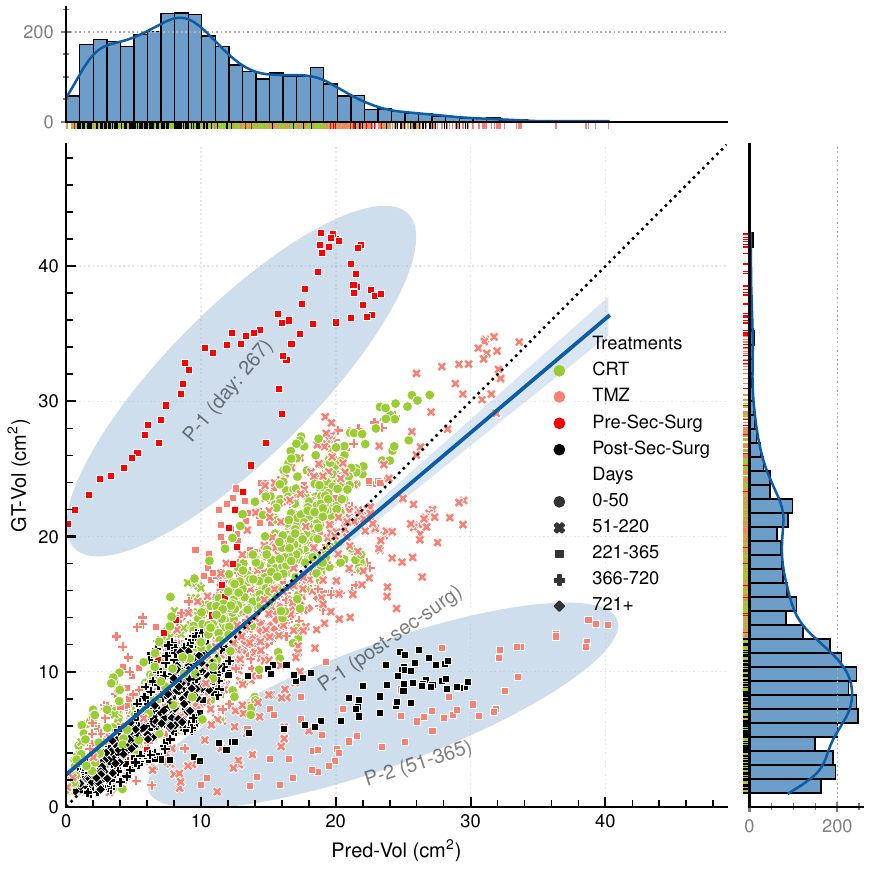}}
    \caption{Tumor growth prediction evaluation based on treatment variables including CRT, TMZ, Second-Surgery (divided into Pre-Sec-Surg and Post-Sec-Surg), as well as their corresponding day ranges. The treatment types are marked by different colors, and the day ranges are indicated by various markers. Two ellipses with annotation texts highlight two outlier regions.}%
    \label{fig:pair_plot2}
\end{figure}
\begin{figure}[!htp]
    \centerline{\includegraphics[width=\columnwidth ]{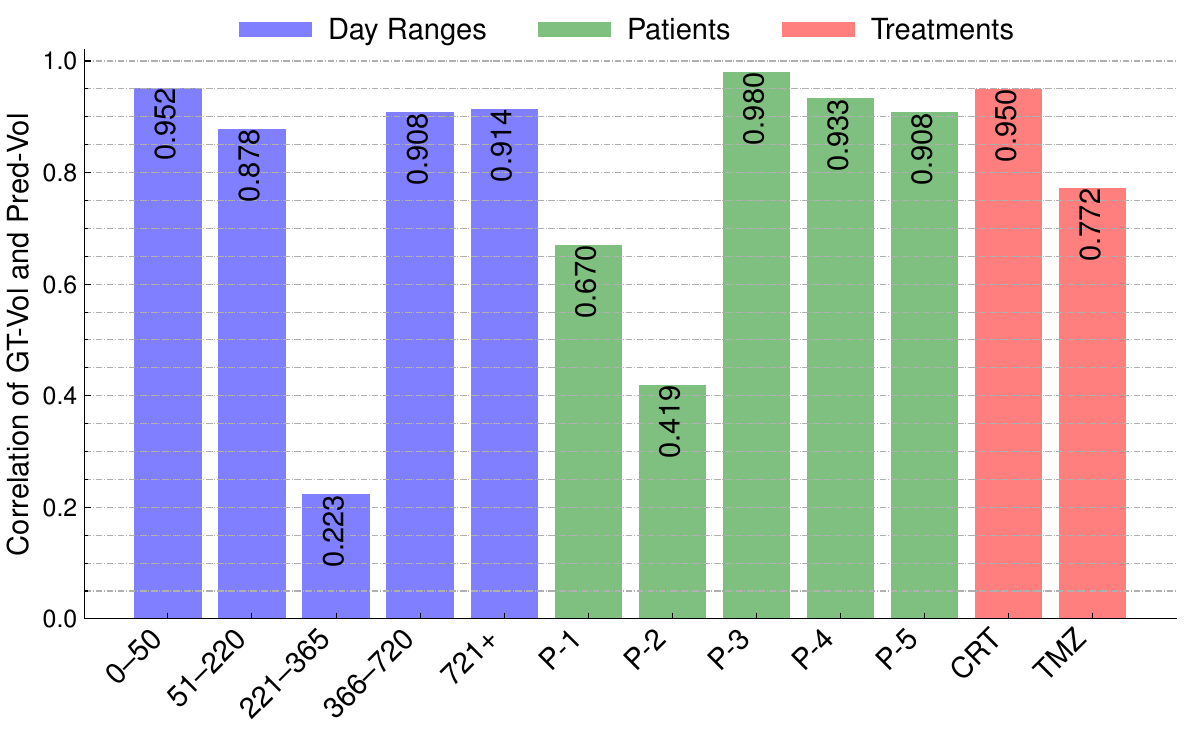}}
    \caption{Pearson correlation coefficients for GT-Vol and Pred-Vol across different day ranges, patients, and treatment types.}%
    \label{fig:correlation}
\end{figure}

We tested our model on the slices containing a tumor of at least $1$ $\text{cm}^2$ at each target session, resulting in a total of $3,352$ local test slices and a total of $8,976$ external test slices. The model requires the three most recent historical MRIs as input. If fewer sequences are available, the most recent MRI is duplicated to meet this requirement. Using the most recent data points (e.g., t2, t3, t4) provides the best results for generating subsequent images (e.g., t5). Although the model can utilize earlier data points (e.g., t1, t2, t3) to predict images at later time points (e.g., t5, t6, etc.), the accuracy and effectiveness of these predictions tend to diminish as the time between sessions increases. Specifically: 1) DSC drops from $0.85$ to $0.46$ as the time interval extends from $0.5$ to $24+$ months, 2) Uncertainty increases progressively, as reflected by the standard deviation widening from $0.09$ to $0.32$, 3) Prediction accuracy remains most reliable within a 4-month window (Mean Dice $0.75 \pm 0.12$). Thus, our model can forecast tumor growth relatively accurately up to 4 months in advance, with diminishing precision beyond this range due to accumulating uncertainties. Notably, predictions for t5 or beyond can still achieve reasonable accuracy when using t1, t2, and/or t3 as inputs without t4, provided the treatment interval between t2/t3 and t5 (or beyond) is less than 4 months.
\subsubsection{Local test results}
In the following two tables, we summarized the results of our model on three tasks: multi-parametric MRI generation, future tumor prediction, and source tumor segmentation, in which $\uparrow/\downarrow$ higher/lower is better. Table~\ref{tab: patient} shows the test performance of each patient (i.e., P-1, P-2, P-3, P-4, and P-5) across various ages ($32-65$ years old) and survival time ($11-83$ months). Table~\ref{tab: treat} summarized the results into five treatment-day ranges, i.e., $0-50$ (about the first 6 weeks of CRT treatment), $51-220$ (around 24 weeks of TMZ treatment in 6 cycles), $221-365$ (one-year survival), $366-720$ (two-year survival), and $721+$ (over two-year survival). Overall, we observed that the model is able to account for stable and progressive tumor cases with growth and shrinkage in different locations. Figure~\ref{results_growth} shows the qualitative comparisons of the predictions from our model and the ground truth on the test set. 

For multi-parametric MRI generation tasks, our model demonstrated high performance with average SSIM: $0.919\pm 0.03$, PSNR: $27.9 \pm 1.2$, and MSE: $0.174 \pm 0.05$. Figure~\ref{violin_plot} compares the generation metric’s distributions regarding each treatment (CRT and TMZ) for each patient. It's worth noticing that TMZ has a long-tail distribution below the first quartile for patients (P-1 and P-2) with SSIM (dashed lines represent the quartiles for each group), whereas CRT has relatively stable distributions across all patients and metrics. In Table~\ref{tab: treat}, we also observed the reduced performance (SSIM: $0.877$) in the treatment-day range of $221-365$. We think this is due to two factors: first, P-1 had a second surgery treatment on day 365, which was beyond the capability bound of our trained model with only two treatment types, and second, P-2 was diagnosed with secondary glioblastomas, which is a fairly uncommon case in our training set, so our model couldn't predict well on such cases.

\begin{figure*}[htp]
\centerline{\includegraphics[width=0.99\textwidth]{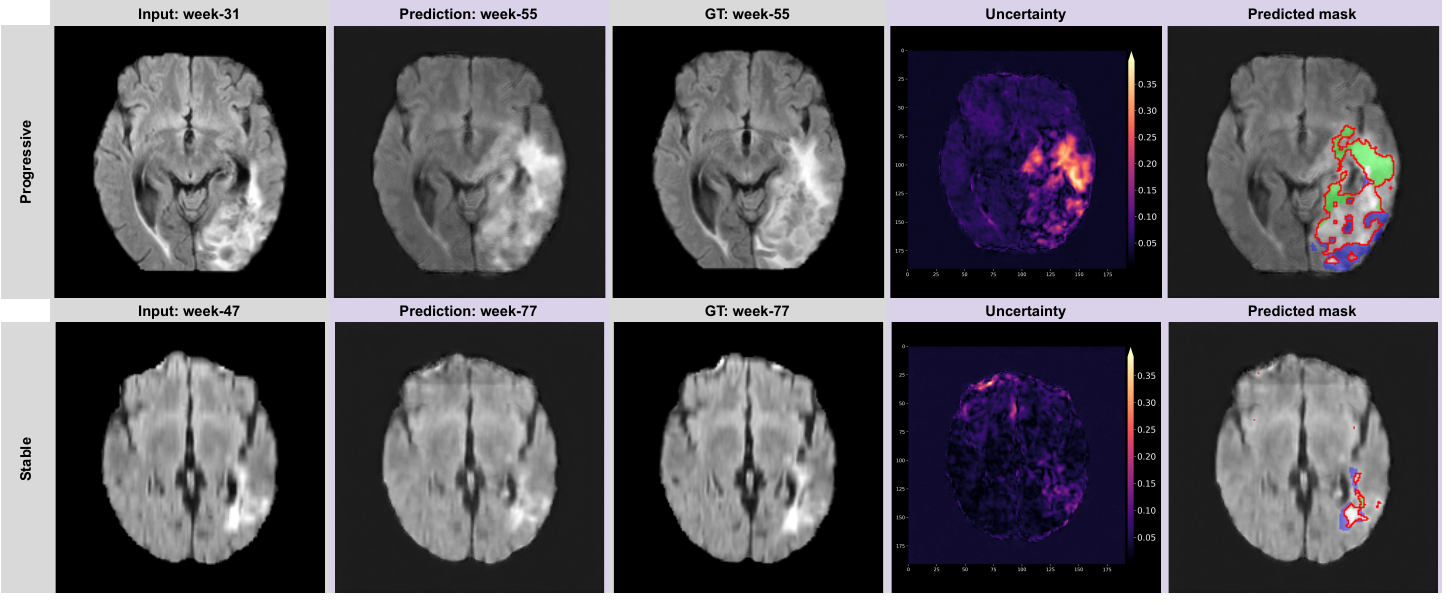}}
\caption{Examples of our model's predictions on the external test dataset. Two representative patient cases: one (the top) with a progressive tumor and another (the bottom) with a stable tumor. In the predicted tumor masks, green areas indicate tumor growth direction, while blue areas indicate tumor shrinkage direction. The ‘Uncertainty’ map uses a heat map to highlight areas of prediction uncertainty on the MR image.}
\label{lumiere_plot}
\end{figure*}


Regarding future tumor prediction and source tumor segmentation tasks, the model achieves an overall DSC of $0.719\pm 0.13$ and $0.849\pm0.09$ respectively. We also noticed that unknown treatment type (e.g. second surgery on P-1) and uncommon glioma type (e.g. secondary glioblastoma of P-2) had a significant impact on the model's prediction of tumor growth. Patients (P-3-4-5) with primary glioblastomas taking standard treatments (CRT/TMZ) obtained better tumor growth prediction than other patients (P-1-2) (ref to Table~\ref{tab: patient}), and the treatment-day range of $221-365$ is the most difficult time spot for predicting tumor progression among all treatment days (ref to Table~\ref{tab: treat}). Figure~\ref{box_plot} illustrates the distributions of RVD and DSC for predicted future tumor and source tumor respectively across different grouped treatment days. Note that outliers of RVD are not shown for simplicity and better visualization reasons, whereas outliers for DSC are shown as marker '$\times$'. We also illustrate pair correlation distributions of GT-Vol (short for the ground-truth future tumor volume on a slice), and Pred-Vol (short for predicted future tumor volume) across all test patients on different treatments and day ranges in Figure~\ref{fig:rel_plot},~\ref{fig:pair_plot1} and~\ref{fig:pair_plot2}. From these figures, it can be seen that:
\begin{itemize}
    \item  GT-Vol and Pred-Vol have a similar distribution with similar mean and variance on each test patient with a strong positive correlation, and the overall slope of the regression line (Figure~\ref{fig:pair_plot1}), is close to 1, indicating that there is a good growth prediction that matches ground-truth in general, whereas Figure~\ref{fig:pair_plot1} and~\ref{fig:pair_plot2} show that P-2 has over-predictions on tumor growth during TMZ treatment days 51-365, and P-1 has significantly more outliers than most other patients, e.g., under-prediction on day 267 (gt-vol greater than pred-vol), and over-prediction on day 365 (pred-vol greater than gt-vol). We also computed Pearson correlation coefficients~\cite{cohen2009pearson} between GT-Vol and Pred-Vol across different day ranges, patients, and treatment types as shown in Figure~\ref{fig:correlation}. These quantitative results highlight strong positive correlations in most day groups, except for the 221–365 group, where the correlation drops significantly due to the increased variability caused by outliers (e.g., patients undergoing second surgery or having secondary glioblastomas). Among individual patients, P-3, P-4, and P-5 demonstrate strong correlations of $0.980$, $0.933$, and $0.908$, respectively, indicating consistent prediction accuracy for these cases. However, P-1 and P-2 exhibit lower correlations of $0.670$ and $0.419$, respectively, primarily due to P-1 undergoing a second surgery and P-2 presenting with a secondary glioblastoma, both of which deviate from the training distribution. The predictions correlate more strongly with ground truth volumes under CRT treatment compared to TMZ. This disparity likely reflects the broader variability in tumor response during TMZ treatment, as highlighted in the following discussion section.
    
    \item The treatment CRT has more points concentrated around the regression line than the TMZ treatment, which has more scattered points with some obvious outliers. Particularly for patients P-1 and P-2, see Figures~\ref{fig:rel_plot} and~\ref{fig:pair_plot1}, there are more outliers in 221-365 and 366-720 day ranges, indicating that the tumor growth prediction becomes worse, which also suggests that the two patients had some abnormal changes or interventions during this period. In this work, we know the possible reason is that, for P-1, the patient had a second surgery on days 267-365 during the TMZ course, whereas for P-2, the patient at the time of diagnosis was classified as secondary glioblastoma. This is why we specifically show an additional treatment variable in Figure~\ref{fig:pair_plot1}, i.e., the second surgery, which is divided into Pre-Sec-Surg (short for pre-second-surgery) and Post-Sec-Surg (short for post-second-surgery). However, our current model was trained only on two treatment types (CRT and TMZ) and assumed there is no second surgery intervention during both CRT and TMZ courses. 
\end{itemize}


\subsubsection{External test results}
We evaluated our model's performance on the external test dataset. Our main focus was assessing the model's capability and generalizability in generating future MRIs, emphasizing metrics like SSIM, PSNR, and MSE. The lack of annotated tumor masks hindered our ability to quantitatively analyze our model's tumor mask predictions, including metrics like DSC and RVD used in our local test. 

Table~\ref{tab: ext_test} presents overall external test results compared to local test results. We noted a minor, yet well-accepted performance decline ($-7.1$\% in SSIM, $-4.6$\% in PSNR, and $+6.8$\% in MSE) in the external test set, which may be linked to disparities in data bias, including variations in image resolution, contrast, and demographic factors. Nonetheless, the model's ability to generalize across datasets indicates promising potential.

Moreover, Figure~\ref{lumiere_plot} illustrates a qualitative assessment of our model's predictions for two distinct tumor scenarios: stable tumors and those showing progressive growth. Our model identifies regions of tumor development while also providing useful uncertainty estimates. These findings enhance our understanding of the model's effectiveness in predicting tumor behavior, thus supporting more informed clinical decision-making processes.

However, we acknowledge the scarcity of high-quality annotated datasets, both local and external, which are essential for validating and refining our model's performance. Currently, we are actively gathering and annotating additional data to facilitate further improvements to our model in the future.
\begin{table}[!htp]\centering
\begin{adjustbox}{max width=\columnwidth}
\begin{threeparttable}
\caption{Performance comparison between local tests and external tests.}\label{tab: ext_test}
\begin{tabular}{lcccccccc}\toprule
\multirow{2}{*}{Test-Data} &\multicolumn{6}{c}{Future MRI Generation}  \\\cmidrule{2-7}
&SSIM $\uparrow$ &$\Delta\%$   &PSNR $\uparrow$  &$\Delta\%$ &MSE $\downarrow$   &$\Delta\%$  \\\cmidrule{1-7}
Local & $0.919 $ & - & $27.97 $ &- & $0.174 $ &-\\
External & $0.848 $ & $-7.1 $ & $27.51 $ &$-4.6$ & $0.242 $ &$+6.8$\\
\bottomrule
\end{tabular}
\end{threeparttable}
\end{adjustbox}
\end{table}

\subsection{Discussion}

\textbf{Effect of the treatment-aware conditioning:} Different treatments can significantly alter the progression patterns of tumors. In predictive modeling of tumor progression, incorporating treatment information as a feature allows for more accurate predictions. In Table~\ref{tab: ablation1}, we evaluated the impact of the treatment-aware conditioning method on our model's performance. We trained a baseline model (baseline-1) without treatment information, achieving an SSIM score of 0.882 for the MRI generation task and a DSC score of 0.556 for the future tumor prediction (segmentation) task. The results show that incorporating treatment-aware input enhances performance. Specifically, the SSIM score increased from 0.882 to 0.919 ($+3.7\%$), indicating a higher quality of generated MRI images when the model is treatment-aware. Similarly, the DSC score rose from 0.556 to 0.719 ($+16.3\%$), demonstrating a notable improvement in the accuracy of future tumor predictions. Incorporating treatment awareness provides valuable insights into the interactions between different treatments and tumor biology. These insights could potentially contribute to more personalized treatment planning.
\begin{table}[!htp]\centering
\begin{adjustbox}{max width=\columnwidth}
\begin{threeparttable}
\caption{The effect of the treatment-aware conditioning method.}\label{tab: ablation1}
\begin{tabular}{cccccccc}\toprule
\multirow{2}{*}{Model}& \multirow{2}{*}{Treatment-aware} &\multicolumn{2}{c}{MRI Generation} &\multicolumn{2}{c}{Tumor Prediction} \\\cmidrule{3-6}
& &SSIM $\uparrow$ &$\Delta\%$ $\uparrow$  &DSC $\uparrow$  &$\Delta\%$ $\uparrow$  \\\cmidrule{1-6}
baseline-1& \xmark & $0.882 $ & - & $0.556 $ &- \\
ours & \cmark & $0.919 $ & $+3.7 $ & $0.719 $ &$+16.3$ \\
\bottomrule
\end{tabular}
\begin{tablenotes}
          \footnotesize   
          \item[*] Note that \cmark and \xmark, indicate whether the model was trained with or without treatment-aware input. 
\end{tablenotes}
\end{threeparttable}
\end{adjustbox}
\end{table}

\textbf{Effect of the joint-task learning: }We also investigated our model's performance by removing key components of the joint-task learning (J-learning) mechanisms: the J-learning of the diffusion probabilistic model with the segmentation network and the Dilated Longitudinal Tumor Fusion weighting (DF-Weighting, Eq.~\ref{weighting}) approach. As shown in Table~\ref{tab: ablation2}, the results clearly show that applying J-learning without DF-Weighting improves performance in both MRI generation (SSIM: +2.1\%) and future tumor prediction tasks (DSc: +13.1\%) compared to the baseline-2 model without J-learning and DF-weighting. However, when both J-learning and DF-Weighting strategies are applied, our model achieves even greater improvements: SSIM: +4.9\% and DSC: +17.8\%. This demonstrates the proposed joint-task learning with the DF weighting mechanism, enabling the model to more effectively capture glioma dynamics from crucial peritumoral tissues in glioma progression.
This approach improves MRI generation quality and enhances tumor growth predictions.
\begin{table}[!htp]\centering
\begin{adjustbox}{max width=\columnwidth}
\begin{threeparttable}
\caption{The effect of the joint-task learning mechanisms.}\label{tab: ablation2}
\begin{tabular}{cccccccc}\toprule
\multirow{2}{*}{Model} &\multirow{2}{*}{J-Learning} &\multirow{2}{*}{DF-Weighting} &\multicolumn{2}{c}{MRI Generation} &\multicolumn{2}{c}{Tumor Prediction} \\\cmidrule{4-7}
& & &SSIM $\uparrow$ &$\Delta\%$ $\uparrow$  &DSC $\uparrow$  &$\Delta\%$ $\uparrow$  \\\cmidrule{1-7}
baseline-2 & \xmark &\xmark & $0.870 $ & - & $0.541 $ & - \\ 
ours & \cmark &\xmark & $0.891 $ & $+2.1 $ & $0.672 $ &$+13.1$ \\
ours & \cmark &\cmark & $0.919 $ & $+4.9 $ & $0.719 $ &$+17.8$ \\
\bottomrule
\end{tabular}
\begin{tablenotes}
          \footnotesize   
          \item[*] Note that the baseline-2 model was trained in two sequential stages. First, it was trained using only the diffusion-branch loss for the future MRI generation task. In the second phase, training focused on tumor growth prediction (segmentation) loss without DF-weighting, with the diffusion branch frozen to isolate and enhance performance on the tumor prediction task.
\end{tablenotes}
\end{threeparttable}
\end{adjustbox}
\end{table}

\textbf{Outliers: }Upon inspection, we noted that the vast majority of poor predictions were concentrated on patients P-1 and P-2 in TMZ treatment day range of $221-365$ as shown in Figure~\ref{fig:rel_plot}. These patients, P-1 (33 years old) and P-2 (44 years old), were relatively young; the glioma growth pattern for young patients might be more uncertain in the $221-365$ day range than for old patients (above 50 years old). For example, P-1's tumor grew at a rapid rate shown in the 267-day MR exam, which exceeded the mean distribution predicted by our model, as shown in Figure~\ref{fig:pair_plot1}. P-2 had secondary glioblastoma~\footnote{According to~\cite{ohgaki2013definition}, secondary glioblastomas progress from low-grade diffuse astrocytoma or anaplastic astrocytoma, which manifest in younger patients, typically with less necrosis, are preferentially located in the frontal lobe, and carry a significantly better prognosis. Despite a similar histologic appearance, primary and secondary glioblastomas are distinct tumor entities that originate from different precursor cells and may require different therapeutic approaches.} rather than primary glioblastoma, which is common in young patients but fairly rare (2 cases) in our training dataset. This helps to explain why the model over-predicted the tumor's growth on P-2 i.e. many Pred-Vol are larger than GT-Vol as shown in Figure~\ref{fig:rel_plot} and~\ref{fig:pair_plot2}. Furthermore, our current model can only predict tumor growth well based on CRT or TMZ treatment with the assumption that there are no other types of interventions (such as second surgery). Because the current model has no way of knowing or estimating which part of the tumor was removed by the second surgery, resulting in our model gives an over-prediction on P-1  i.e. Pred-Vol is well larger than GT-Vol after P-1 underwent a second surgery (Post-Sec-Surg) shown in Figure~\ref{fig:rel_plot} and~\ref{fig:pair_plot1}. To improve the handling of outliers and special cases, such as secondary glioblastomas or post-second surgery scenarios, the best approach is to increase the number of relevant samples in our dataset. Currently, our model's training data includes very few of these cases, which limits its predictive capabilities in these scenarios. By incorporating more examples of these special cases in future datasets and further fine-tuning our model with more than three treatments (such as CRT, TMZ, Sec-Surgery, etc.), we anticipate significant improvements in its predictive performance for these situations.

\textbf{Aid treatment planning:}
Our model's ability to predict future tumor growth can potentially aid clinical workflows. For instance, if a patient's MRI data from past exams shows a certain progression trend, our model can forecast tumor growth over the next few months. For example, if the prediction indicates a rapid growth then it might prompt the clinician to re-evaluate the treatment plan and possibly order the next MRI scan sooner. This predictive capability can aid in earlier intervention strategies, optimize treatment schedules, and potentially improve patient prognoses. By providing accurate forecasts of tumor development, our model can help clinicians make more informed decisions, tailor treatments to individual patients, and allocate resources more effectively, with the potential goal of improving patient outcomes.

\textbf{Interpreting uncertainty maps:}
In our model, uncertainty estimates are quantified using probabilistic methods that capture variance in predictions. Specifically, the uncertainty quantified primarily reflects data uncertainty, which arises from the inherent variability in the input data and the stochastic nature of the sampling process during inference. These uncertainty maps highlight areas where the model's predictions are less certain, aiding clinicians in treatment planning. Uncertainty maps can identify tumor regions where predictions are less confident, helping assess the risk of progression and informing treatment decisions. In high-uncertainty areas, clinicians might choose a conservative approach initially, followed by closer monitoring and adaptive strategies. For instance, if uncertainty is high in predicting tumor boundaries, combining surgical resection with radiation or chemotherapy might ensure comprehensive treatment. While our current approach focuses on data uncertainty derived from standard deviation calculations across multiple samplings, incorporating a Bayesian framework~\cite{kendall2017uncertainties} could provide additional insights, particularly into model uncertainty, which arises from the parameters of the model itself. Bayesian methods, such as Monte Carlo Dropout~\cite{gal2016dropout} or Bayesian Neural Networks~\cite{hernandez2015probabilistic}, offer mechanisms to quantify both data and model uncertainty systematically. However, implementing a Bayesian framework introduces significant computational overhead and complexity, which can be challenging in resource-constrained clinical environments. Future work will explore the feasibility and trade-offs of such Bayesian methods to enhance uncertainty quantification and provide a more comprehensive assessment of predictive reliability.

\textbf{Generalization:} 
Our methodology indeed primarily focuses on the treatment-aware Diffusion (TaDiff) probabilistic model for glioma prediction. While our current discussion predominantly revolves around its application to diffuse gliomas, it's important to clarify that our model's framework in theory may not inherently be restricted to this specific disease. The model's adaptability to predict the evolution of various solid tumors relies on the availability of comprehensive longitudinal data that includes both imaging and treatment information. However, it is crucial to note that the model's performance depends on the ability to normalize longitudinal images within a consistent spatial framework. While this is relatively straightforward for brain lesions due to the anatomical constraints of the skull, it presents a greater challenge for other tumors, such as those in the breast, due to the less restrictive and more variable anatomical environment. The progression of breast cancer, like many other solid tumors, often involves similar dynamics of tumor growth and response to treatment over time. Our model’s principles, rooted in analyzing longitudinal imaging and treatment data, can be seamlessly applied to predict the evolution of breast cancer tumors or other solid tumor types. However, it’s crucial to emphasize that the adaptability of our model relies heavily on the availability of comprehensive time-series data encompassing both imaging and treatment information for the specific tumor type under consideration. Regrettably, the current landscape of publicly available and high-quality longitudinal imaging and clinical data is sparse, which inherently limits researchers’ ability to explore this field extensively.

\textbf{Limitations and future work: }Our work has some limitations that are discussed herein. First, as a data-driven deep-learning-based method, the model requires more data than methods based on explicit biological diffusion models. The current sample is not sufficient to capture all possible tumor growth or shrinkage patterns related to various diffuse glioma types (such as primary or secondary glioblastomas) and treatment protocols (such as CRT, and TMZ with or without second surgery), and the test set of 5 subjects is relatively small; hence a larger sample is needed for more extensive evaluation. This is also one of the main reasons why we chose to work on 2D models rather than 3D models, as the latter would significantly reduce the number of available training and test samples. Second, it is important to note the high computational cost and slow inference/sampling process of DDPM models, which need large batch sizes and hundreds of diffusion steps. To speed up the inference process, future work could be directed to improve sampling methods such as DPM-Solver~\cite{lu2022dpm}. This method can potentially improve sampling speeds by 20–30 times without compromising the quality of the generated samples. We prioritize this optimization in our future study. Third, hybrid deep learning methods that integrate computational models~\cite{gholami2016inverse} offer valuable biophysical insights for tumor growth prediction, particularly in data-limited scenarios. Recent approaches, such as TGM-Nets~\cite{chen2023tgm} and DL-PDE~\cite{weidner2024learnable}, have demonstrated promising performance in long-term predictions and biological interpretability on simulated tumor datasets. However, these methods lack the treatment-aware and comprehensive predictive capabilities of our model. Nevertheless, their complementary strengths suggest promising avenues for future research, particularly in the integration of computational tumor modeling with our TaDiff approach.
Additionally, while our results are promising, the clinical utility of our model needs to be further fine-tuned and validated on larger and more diverse datasets from multiple sites. We are confident that larger datasets will become available in the future, which will enable us to further improve the robustness and accuracy of our method.

\section{Conclusions}\label{concl}
We present a treatment-aware diffusion (TaDiff) model for generating longitudinal multi-parametric MRI and predicting diffuse glioma tumor growth. The proposed method incorporates a sequence of source MRI and treatment information as conditioning inputs and employs an optimized joint learning strategy that effectively leverages the benefit of both DDPM and segmentation networks, allowing for the generation of future MRI for any given treatment and time point, as well as the prediction of corresponding glioma growth probabilities. Our model also enables the ensemble of multiple predictions using the stochastic sampling process to account for various possible treatment responses and further produces uncertainty maps by calculating the variance of these predictions. This approach has the potential to predict treatment-aware tumor growth patterns and quantify tumor evolution uncertainty, thereby supporting improved treatment planning and, ultimately, enhancing patient outcomes.

\appendices

\section*{Acknowledgment}
The authors want to thank the Digital Research Alliance of Canada for providing free computing resources, as well as CRUE-Universitat Politècnica de València for funding the open access charge. The authors would also like to thank all of the reviewers (whether anonymous or not) for their insightful comments and careful reading of the manuscript.

\bibliographystyle{ieeetr}
\bibliography{main}

\end{document}